\documentclass[twocolumn,preprintnumbers,amsmath,amssymb]{revtex4-1}

\usepackage{graphicx}
\usepackage{dcolumn}
\usepackage{bm,epsfig}
\usepackage{epstopdf}
\usepackage{amsmath}
\usepackage{amsfonts}
\usepackage{amssymb}

\begin{document}

\title{Narrow-gap semiconducting behavior in antiferromagnetic Eu$_{11}$InSb$_9$}
\author{S. S. Fender, S. M. Thomas, F. Ronning, E. D. Bauer, J. D. Thompson, and P. F. S. Rosa}
\affiliation{Los Alamos National Laboratory, Los Alamos, New Mexico 87545, U.S.A.}
\date{\today}

\begin{abstract}
Here we investigate the thermodynamic and electronic properties of Eu$_{11}$InSb$_9$ single crystals. Electrical transport data show that Eu$_{11}$InSb$_9$ has a semiconducting ground state with a relatively narrow band gap of $320$~meV. Magnetic susceptibility data reveal antiferromagnetic order at low temperatures, whereas ferromagnetic interactions dominate at high temperature. Specific heat, magnetic susceptibility, and electrical resistivity measurements reveal three phase transitions at 
$T_{N1}=9.3$~K, $T_{N2} =8.3$~K, and $T_{N3} =4.3$~K. Unlike Eu$_{5}$In$_{2}$Sb$_6$, a related europium-containing Zintl compound, no colossal magnetoresistance (CMR) is observed in Eu$_{11}$InSb$_9$. We attribute the absence of CMR to the smaller carrier density and the larger distance between Eu ions and In-Sb polyhedra in Eu$_{11}$InSb$_9$. Our results indicate that  Eu$_{11}$InSb$_9$ has potential applications as a thermoelectric material through doping or as a long-wavelength detector due to its narrow gap. 
\end{abstract}

\maketitle

\section{INTRODUCTION}
Narrow-gap semiconductors have drawn renewed interest due to a number of promising applications, which include thermoelectric devices and 
photoelectrodes, dark matter detectors, and quantum computing platforms based on topological properties \cite{first,photoelectrode,second,darkmatter,topins}.
These diverse applications rely on the discovery of new, stable, and clean candidates. Zintl-phase materials are a particularly promising route in 
the design of narrow-gap semiconductors owing to their inherent charge balance. Zintl intermetallic phases are a combination of alkaline, alkaline-earth, or rare-earth cations and a covalently bonded polyanionic structure that achieves an octet in the valence electron shell \cite{zintldef}. Zintl compounds are valence precise, and electron transfer between cations and anions is expected to create a semiconducting state~\cite{zintl}. Further, the inclusion of rare-earth 
elements with localized $f$ electrons may enable magnetic order and reduce sensitivity to air and moisture compared to their alkaline and alkaline-earth analogues \cite{Rosa, Yb11Eu11,Eu5}.

The large number of narrow-gap semiconductors reported within the Zintl concept points to a promising pathway to realizing new narrow-gap materials. For instance, investigations of Yb$_{11}$GaSb$_9$, Yb$_{11}$InSb$_9$, Yb$_{11}$AlSb$_9$, Ca$_{11}$AlSb$_9$, and Sr$_{11}$AlSb$_9$ support the notion that the Zintl formalism gives rise to a narrow-gap semiconducting state \cite{Ga,Al,Ca}. 
Notably, Yb$_{11}$GaSb$_9$ displays semiconducting behavior in electrical resistivity, but the presence of significant sample dependence indicates that impurity states play a role, e.g., small amounts of Yb$^{3+}$ from Yb$_{2}$O$_{3}$ \cite{bobev}. Conversely, isostructural Eu$_{11}$InSb$_{9}$ was reported to be paramagnetic above 8~K, which is unusual given the expected divalent europium configuration, and its electrical resistance showed a weakly metallic behavior, which marks a distinct deviation from the Zintl formalism in this family \cite{Yb11Eu11}.

Useful insights can be obtained by comparison with structurally related compounds. For example, Eu$_5$In$_2$Sb$_6$ and Yb$_5$In$_2$Sb$_6$ are narrow-gap semiconductors that satisfy the Zintl formalism and show promise as thermoelectric materials \cite{Eu5,Yb5}. Eu$_5$In$_2$Sb$_6$ also undergoes a dramatic reduction in electrical resistance in the presence of magnetic fields \cite{Rosa}. This property, initially observed in perovskite manganites, is coined colossal magnetoresistance (CMR) \cite{nature} and has been attributed to the presence of magnetic polarons, $i.e.$, quasiparticles arising from strong exchange coupling between free carriers and background spins of divalent europium. The large negative magnetoresistance in Eu$_5$In$_2$Sb$_6$ is among the largest CMR observed to date, which invites further investigation of Eu-containing narrow-gap materials~\cite{EuB6, EuTe2, EuO, EuIn2As2, EuP3}.

\begin{figure*}[!ht]
      \begin{center}
      \includegraphics[width=2\columnwidth]{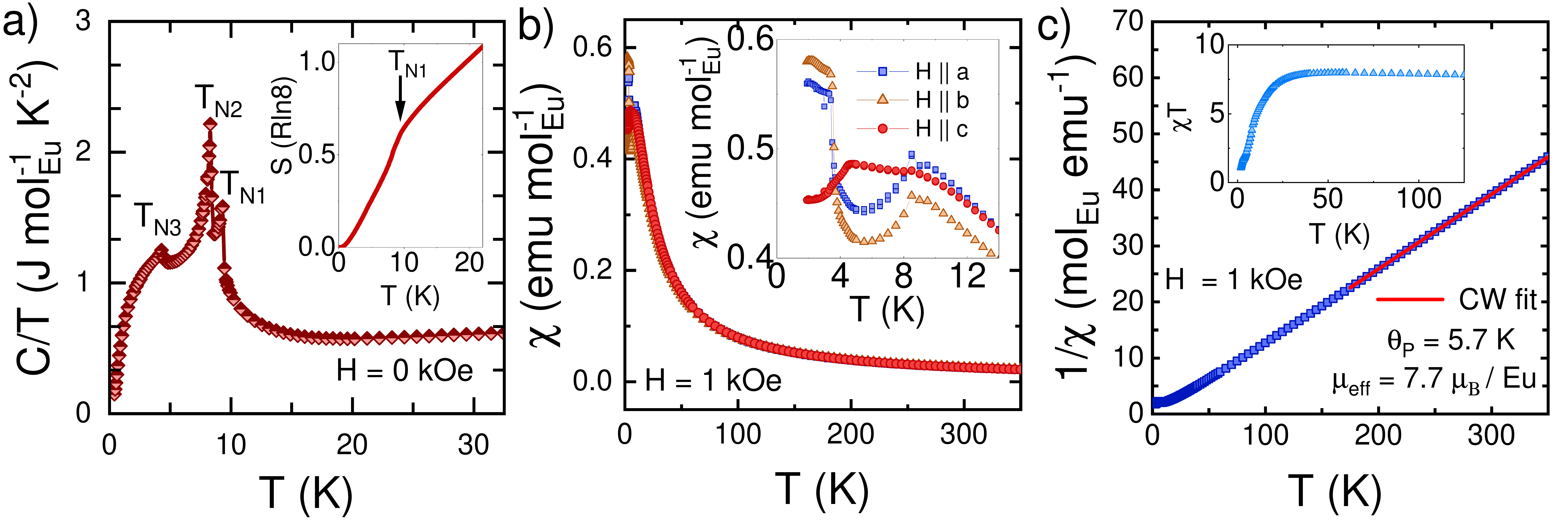}
      \end{center}
      \vspace{-0.55cm}
      \caption{a) Specific heat divided by temperature T, C/T vs. T, at zero magnetic field of Eu$_{11}$InSb$_9$. The inset displays the entropy in units of $R$ln8.  b)~Temperature dependent magnetic susceptibility $\chi$(T) with a field of 1~kOe applied along distinct crystallographic directions. The inset shows the low-temperature anisotropic $\chi$(T) data. c) Inverse magnetic susceptibility 1/$\chi$(T) $vs.$ T. A Curie-Weiss fit is shown by a solid line. The inset shows $\chi$T as a function of temperature.} 
      \label{fig:Fig1}

      \end{figure*}

Here we investigate the structural and physical properties of Eu$_{11}$InSb$_9$ single crystals grown by the self-flux method. Eu$_{11}$InSb$_9$ crystallizes in an orthorhombic crystal structure displaying nonsymmorphic symmetries but no inversion symmetry (space group 45, $Iba2$). In this structure, six inequivalent europium sites are located between polyanionic 
InSb$_4$ tetrahedra, Sb$_2$ dimers, and Sb ions. To achieve charge balance, europium ions are expected to be divalent in line with [11Eu$^{2+}$][InSb$_4^{9-}$][Sb$_2^{4-}$][3(Sb$^{3-}$)]. Because Eu$^{2+}$ ions have a large local moment, this electron count indicates that Eu$_{11}$InSb$_9$ will likely order magnetically.
In fact, our magnetic susceptibility and specific heat results show that Eu$_{11}$InSb$_9$ has a complex magnetic ground state and undergoes three magnetic phase transitions at $T_{N1}$= 9.3~K, $T_{N2}$= 8.3~K, and $T_{N3}$= 4.3~K. Both the entropy recovered at the magnetic transitions and the high-temperature Weiss temperature are consistent with divalent europium. In addition, our electrical resistivity data display a relatively small activation energy of 160~meV and low carrier density of $10^{15}$ holes/cm$^{3}$, which are consistent with narrow-gap semiconducting behavior. At low temperatures, however, a departure from the activated behavior points to the presence of in-gap conducting channels. Compared to Eu$_5$In$_2$Sb$_6$, the coupling between $f$ and conduction electrons in Eu$_{11}$InSb$_9$ appears to be much smaller, and colossal magnetoresistance is not observed. We discuss possible origins for the small magnetoresistance of Eu$_{11}$InSb$_9$. Our results show that 
Eu$_{11}$InSb$_9$ is a magnetically ordered Zintl semiconductor and invite further investigation of its thermoelectric and long-wavelength detector properties.

\section{Experimental Details}

Single crystals of Eu$_{11}$InSb$_9$ were grown through a self-flux of constituent elements in excess indium with an elemental precursor stoichiometry of 10Eu:60In:8Sb.
The elements were placed in an alumina crucible with a quartz wool filter and then sealed in a silica ampule under partial argon pressure. The ampoule was heated to 1050$^{\circ}$C over the course of 14 hours, held at constant temperature for 24~h, and then cooled to 750$^{\circ}$C at 4$^{\circ}$C/h. 
The flux was subsequently removed by centrifugation.

The crystallographic structure of Eu$_{11}$InSb$_9$ was determined at room temperature by a Bruker D8 Venture single-crystal diffractometer equipped with Mo radiation. Diffraction analysis shows that Eu$_{11}$InSb$_9$ crystallizes in the orthorhombic space group $Iba2$ with lattice parameters $a = 12.24 $ \AA , 
$b = 12.87$~\AA , and $c = 17.32$~\AA. These values match previous structural reports \cite{Yb11Eu11}. Elemental analysis using energy-dispersive x-ray spectroscopy in a FEI Quanta scanning electron microscope resulted in Eu$_{11}$In$_{1.3(2)}$Sb$_{8.9(2)}$. 
Single crystals of Eu$_{11}$InSb$_9$ are significantly less sensitive to air and moisture than the alkaline-earth counterparts, but they were kept in an argon glovebox between measurements to allow for sample stability over several months.  Magnetization measurements were obtained through a Quantum Design SQUID-based magnetometer with 10$^{-8}$ emu sensitivity. 
Specific heat measurements were made using a Quantum Design calorimeter that utilizes a quasi-adiabatic thermal relaxation technique. The electrical resistivity ($\rho$) was measured using a standard four-probe configuration. Due to the poor conductivity of the crystals, spot welding was not possible. 
Pt wires were attached to sputtered gold pads using silver paint. At high temperatures, $\rho$ was measured with a Lakeshore 372 AC bridge, whereas at low temperatures a two-point DC method was required due to the large resistance of the sample.

\section{Results}

First, we devote our attention to the thermodynamic characterization of Eu$_{11}$InSb$_9$. Zero-field heat capacity as a function of temperature is shown in Figure \ref{fig:Fig1}a. Three transitions are identified at T$_{N1}$ = 9.3~K, T$_{N2}$ = 8.3 K, and T$_{N3}$ = 4.3 K. As shown in the inset of Figure~\ref{fig:Fig1}, the entropy recovered at T$_{N1}$ is only 60$\%$ of the expected entropy for the $J=S=7/2$ divalent europium multiplet, and the full $R$ln8 entropy is only recovered at $\sim 20$~K. A similar entropy reduction has been observed
in Zintl antiferromagnet Eu$_3$Sn$_2$P$_4$, which also hosts divalent europium \cite{Eu3Sn2P4}.
The missing entropy at T$_{N1}$ could be either caused by the presence of short-range interactions or by non-ordered europium ions. Though the subtraction of a nonmagnetic phonon background is required for a more quantitative analysis, we note that the missing entropy of 40$\%$ of $R$ln8 is consistent with 4 out of 11 Eu ions not participating in magnetic order. 

\begin{figure*}[!ht]
\begin{center}
\includegraphics[width=1.75\columnwidth]{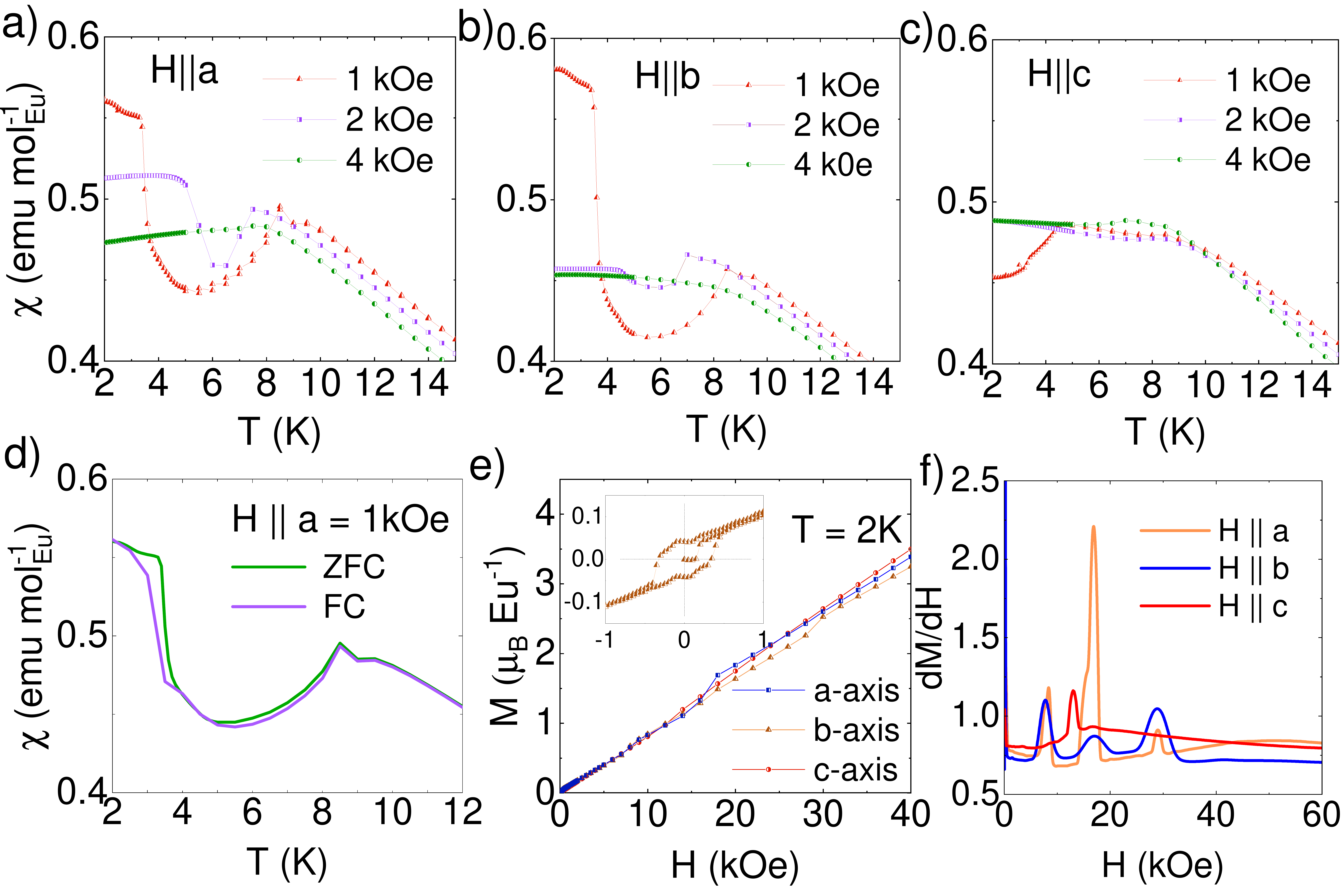}
\end{center}
\vspace{-0.55cm}
\caption{a) Low temperature susceptibility in the a-axis. b) Low temperature susceptibility in the b-axis. c) Low temperature susceptibility in the c-axis. d) Zero-field and field-cooled magnetization curves. e) Low temperature anisotropic magnetization, zero-field cooled, with the inset showing the low-temperature region with hysteresis for $H||b$ around H=0. f) Derivative with respect to field taken from Figure \ref{fig:Fig2}e shows field induced transitions in the a,b and c axis.} 
\label{fig:Fig2}
\end{figure*}

	Figure \ref{fig:Fig1}b shows the magnetic susceptibility, $\chi$=M/H, of Eu$_{11}$InSb$_9$ as a function of temperature. As expected for Eu$^{2+}$, the susceptibility is isotropic at high temperature, whereas the inset of Fig. \ref{fig:Fig1}b highlights the anisotropic behavior of $\chi$(T) in the vicinity of magnetic order. Kinks in $\chi$(T) at $T_{N1}$=9.4~K,  $T_{N2}$= 8.4~K, and  $T_{N3}$= 4.0~K reflect phase transitions, in good agreement with the transition temperatures identified in specific heat data. The inverse susceptibility 1/$\chi$ as a function of temperature follows a Curie-Weiss law at high temperature (T$>$175K), the results of which are shown in Figure \ref{fig:Fig1}c. The linear fit yields a positive Weiss temperature $\theta =$ 5.7 K, which  indicates the presence of ferromagnetic interactions. At low temperatures (T$<$30K), the deviation from Curie-Weiss behavior reflects the onset of short-ranged antiferromagnetic correlations among magnetic moments. The presence of both ferro- and antiferromagnetic interactions, as well as the presence of six inequivalent Eu sites, gives rise to a complex magnetic network with multiple exchange parameters as also observed in Eu$_5$In$_2$Sb$_6$ \cite{Rosa}. From the high-temperature Curie-Weiss fit, the effective moment in Eu$_{11}$InSb$_9$ is 7.7~$\mu_{B}/$Eu, which is within 3$\%$ of 7.94 $\mu_{B}$ expected for a Hund's-rule divalent europium ion. Thus, any crystal-field and Kondo effects are negligible, and the deviation at 30~K can be attributed solely to the onset of antiferromagnetic short-range interactions. Further evidence for the onset of short-range interactions is given by the sharp decrease in $\chi$T below $T^{*}=30$~K (inset of Fig.~1c).

	Magnetic susceptibility and magnetization (M) measurements provide further insight into the magnetic state of Eu$_{11}$InSb$_9$.  The magnetic transitions found in Figures \ref{fig:Fig2}a-c are suppressed with increasing field applied along each of the three axes, once again indicative of antiferromagnetic character. Magnetic hysteresis is not evident between zero-field and field-cooled measurements at T$_{N1}$ and T$_{N2}$ for field applied along the $a$ axis, but weak hysteresis is present at T$_{N3}$ (see Figure \ref{fig:Fig2}d). Consequently, a small hysteresis is also observed in a magnetization loop at 2~K, shown in the inset of Figure \ref{fig:Fig2}e. At a minimum, these results indicate the absence of spin-glass or hard ferromagnetic order. More generally, the temperature dependence of 
	the magnetic susceptibility for $H||a,b$ upon approaching T$_{N3}$ from above and its saturation below T$_{N3}$ reflect a clear change in the nature of magnetic order. Ferrimagnetic or spin-canted order are possible candidates. Additionally, the flattening of low-temperature susceptibility data  below T$_{N1}$ and T$_{N2}$ initially suggests that the c-axis is the hard magnetization axis, but field-induced transitions 
	along all three axes are observed in $dM/dH$ plots shown Figure \ref{fig:Fig2}f. Given the presence of six inequivalent Eu sites, it is not surprising to find such complex magnetic order, which hinders magnetic structure determination from magnetization data alone. Magnetic x-ray diffraction measurements would be valuable to determine the complex magnetic structure of Eu$_{11}$InSb$_9$.

\begin{figure*}[!ht]
\begin{center}
\includegraphics[width=2\columnwidth]{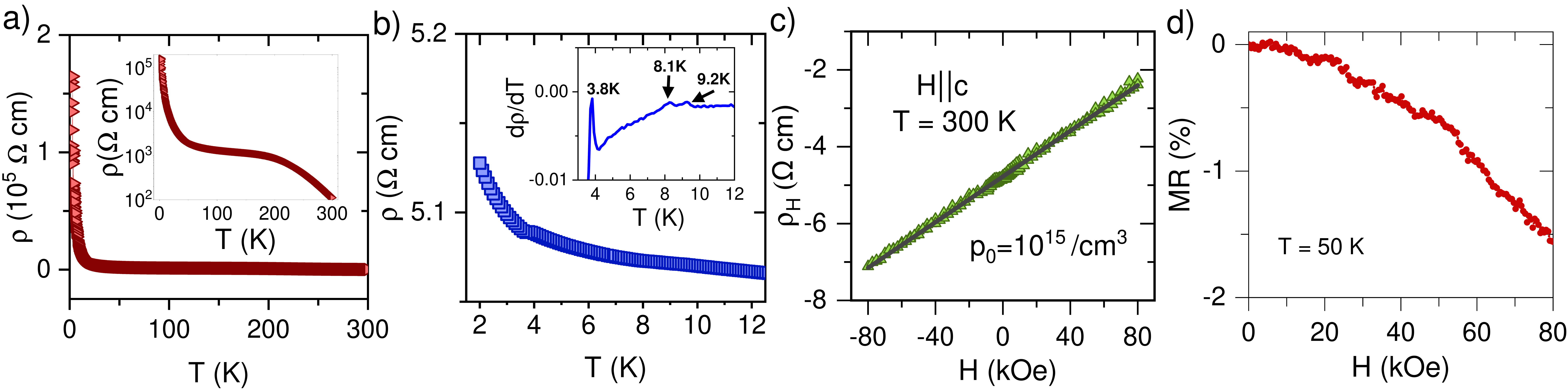}
\end{center}
\vspace{-0.55cm}
\caption{a) Electrical resistivity, $\rho$, versus temperature of Eu$_{11}$InSb$_9$ (sample 1). Inset shows log($\rho$) $vs.$ T. b) Low-temperature electrical resistivity measurements for sample 2. Inset maps transitions identified in both magnetic and specific heat measurements.  c) Hall resistivity $\rho_H$ \textit{vs.} magnetic field of sample 1 measured with field along the $c$ axis. The black line is a linear fit to the data, which agrees with anti-symmetrized data and suggests no measurable contribution from longitudinal resistance. d) Normalized magnetoresistance (MR=($\rho$(T)-$\rho$$_0$)/$\rho$$_0$) as percent for sample 1.} 
\label{fig:Fig3}
\vspace{-0.2cm}
\end{figure*}

        Turning to the electrical properties, Figure \ref{fig:Fig3}a shows the temperature dependent electrical resistivity, $\rho$(T), of Eu$_{11}$InSb$_9$ as measured with current applied in the ab plane. The marked increase in resistivity on cooling is in agreement with the expected semiconducting behavior. The inset of Figure \ref{fig:Fig3}a shows $\rho$(T) in a logarithmic scale to better highlight the presence of distinct resistivity regions, which points to multiple conduction mechanisms. In fact, narrow-gap semiconductors are typically susceptible to defects and impurities, and we find that the electrical resistivity in Figure \ref{fig:Fig3}a is in stark contrast to the low-temperature resistivity data for a second sample (s2) shown in Figure \ref{fig:Fig3}b. This sample is remarkably less insulating than sample s1 and indicates the presence of disorder, defects, or residual flux within the crystals. This result also provides a sensible explanation for the weakly metallic behavior in electrical resistivity observed previously in Eu$_{11}$InSb$_9$ as measured through a pressed pellet sample that may have contained residual flux~\cite{Yb11Eu11}. In addition, we cannot rule out the presence of small amounts of hydrogen in these samples, which would be invisible to x-rays. The effects of hydrogen in the chemistry of $A_{5}Pn_{3}$ ($A=$ alkaline-earth and divalent rare-earth elements, $Pn =$ As, Sb, Bi) has been evaluated in Ref. \cite{Corbett}.
        Notably, the small change in resistivity on cooling through the magnetic transitions suggests a rather reduced coupling between local moments and conduction electrons. Small anomalies taken from $d\rho/dT$, shown in the inset of Figure \ref{fig:Fig3}b, align with those identified from low-temperature magnetic susceptibility and specific heat data. 
                
        Field-dependent Hall resistivity data at 300~K, shown in Figure \ref{fig:Fig3}c, provide valuable information about carrier type and scattering mechanisms in Eu$_{11}$InSb$_9$. The positive Hall coefficient value, 
         $\rho_H$/H = R$_H=$ 3 $\times$ 10$^{3}$ cm$^3$/C, indicates electronic transport dominated by holes. The carrier density is only 10$^{15}$ cm$^{-3}$, as determined by the single-band expression $1/R_{H}e$, and is consistent with most narrow-gap semiconductors \cite{narrowgap}. Notably, Eu$_{11}$InSb$_9$ has a carrier density two orders of magnitude lower than that of Eu$_5$InSb$_{6}$, which was determined to be 10$^{17}$ cm$^{-3}$ at room temperature \cite{Rosa}. The magnetoresistance of Eu$_{11}$InSb$_9$, shown in Figure \ref{fig:Fig3}d, is also six orders of magnitude smaller than that observed in Eu$_5$In$_2$Sb$_6$.

\begin{figure}[!h]
\begin{center}
\includegraphics[width=1\columnwidth]{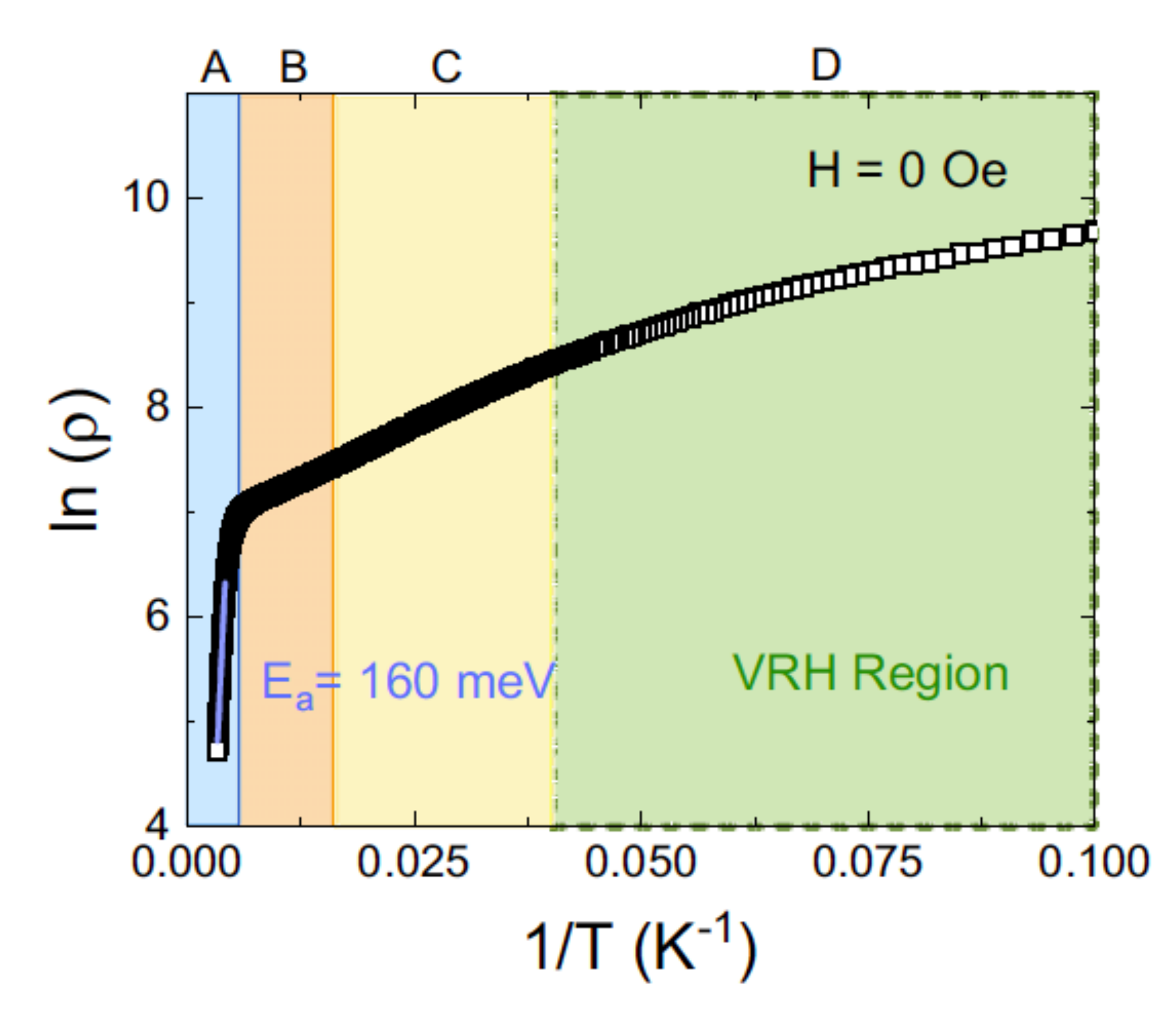}
\end{center}
\vspace{-0.55cm}
\caption{Arrhenius plot (ln($\rho$) $vs.$ 1/T) for Eu$_{11}$InSb$_9$. Resistivity data are from Fig. \ref{fig:Fig3}a.} 
\label{fig:Fig4}

\end{figure}

Figure \ref{fig:Fig4} displays an Arrhenius plot, ln$\rho$ $vs.$ 1/T, wherein a linear fit of the high-temperature linear region ($>$240 K) yields an activation energy of 160~meV, indicative of a relatively narrow gap of 320~meV. In agreement with the inset of Figure \ref{fig:Fig3}a, distinct regions can be identified, and a simple activation fit alone is inadequate for the characterization of the electrical transport at lower temperatures. Narrow-gap semiconductors are typically susceptible to defects and impurities, and the temperature dependence shown in Figure~4  is consistent with that of a lightly doped semiconductor, $i.e.$, region A exhibits an intrinsic conduction range, whereas regions B, C, and D are consistent with a saturation range of impurity conduction, a freeze-out range, and a hopping conduction regime, respectively \cite{springer}.
  
   \section{Discussion}     
The discrepancy between the physical properties of Eu$_{11}$InSb$_9$ and its counterpart Eu$_5$In$_2$Sb$_6$ invites a more in-depth comparison between them. Both compounds are Zintl, nonsymmorphic, antiferromagnetic, narrow-gap semiconductors that share the same constituent elements. Thus, one might expect
 that they share other notable phenomena as well. In particular, colossal magnetoresistance appears to be ubiquitous in low-carrier density Eu$^{2+}$ compounds, 
 such as EuB$_{6}$, EuTe$_{2}$, EuO, and EuIn$_{2}$As$_{2}$, to name a few \cite{EuB6,EuTe2,EuO,EuIn2As2}. This feature, however, is virtually absent in the work presented here for Eu$_{11}$InSb$_9$. 
 One sensible scenario to explain CMR in this class of materials is the presence of magnetic polarons.  For instance, evidence for magnetic polarons in Eu$_5$In$_2$Sb$_6$ comes from a collection of experimental results. First, a deviation from the Curie-Weiss law at temperatures much higher than $T_{N}$ resembles the formation of magnetic polarons in the manganites and is accompanied by the onset of colossal magnetoresistance. Magnetic susceptibility measurements also reveal a ferromagnetic Weiss temperature at high temperatures, which in consistent with the formation of ferromagnetic clusters. Second, at low temperatures a Schottky anomaly in heat capacity measurements combined with a marked decrease in $\chi T$ above $T_{N}$ indicate the presence of short-range inter-polaron antiferromagnetic interactions. Finally, microscopic electron spin resonance measurements reveal a linewidth narrowing at high fields, which is consistent with the polaron picture \cite{Rosa}.
 
Though Eu$_{11}$InSb$_9$ also displays complex exchange interactions that lead to a ferromagnetic Weiss temperature at high temperatures and multiple antiferromagnetic transitions at low temperatures, no evidence for magnetic polarons is observed. The formation of magnetic polarons is largely contingent on two properties: low carrier densities and relatively strong exchange coupling between europium and conduction electron spins \cite{Kaminski}. We argue that there must be a lower limit for the carrier density below which magnetic polaron formation is impaired. In fact, the carrier density of Eu$_{11}$InSb$_9$ is significantly
lower than that in Eu$_5$In$_2$Sb$_6$ and other europium-based CMR materials, which suggests that the formation of ferromagnetic clusters in Eu$_{11}$InSb$_9$ may be hindered by the lack of conduction electrons to self-trap around Eu sites. 
In addition, a comparison between the crystal structures of Eu$_{11}$InSb$_9$ and Eu$_5$In$_2$Sb$_6$ reveals structural elements that provide further explanation for smaller exchange
 interactions in Eu$_{11}$InSb$_9$. Measuring bond distances between Eu atomic centers and the In-Sb polygons reveals that Eu and In-Sb polygons in Eu$_{11}$InSb$_9$ are, on average, 0.2 \AA$\:$  farther apart than those in Eu$_5$In$_2$Sb$_6$. Decreased overlap between \textit{f} electrons of Eu$^{2+}$ and conduction electrons of the In-Sb polygons could lead to smaller exchange coupling and therefore smaller magnetoresistance.

\section{CONCLUSIONS}
Here we report the synthesis of Eu$_{11}$InSb$_9$ single crystals and their characterization through thermodynamic and electrical transport measurements. Our results reveal that Eu$_{11}$InSb$_9$ exhibits
a semiconducting ground state with a relatively narrow gap of 320~meV. Magnetic data show that ferromagnetic interactions dominate at high temperature whereas antiferromagnetic correlations set in at low temperatures ($<$30K). Unexpectedly, Eu$_{11}$InSb$_9$ showed significantly lower magnetoresistance compared to other related 
europium-containing compounds, and we attribute this difference to the smaller carrier densities and exchange couplings in Eu$_{11}$InSb$_9$.
 Our results indicate that  Eu$_{11}$InSb$_9$ has potential applications as a thermoelectric material through doping or as a long-wavelength detector due to its narrow gap.\\ 

\begin{acknowledgments}
Work at Los Alamos National Laboratory (LANL) was performed under the auspices of the U.S. Department of Energy, Office of Basic Energy Sciences, Division of Materials 
Science and Engineering. Scanning electron microscope and energy dispersive X-ray measurements were performed at the Center for Integrated Nanotechnologies, an Office of Science User Facility operated for the U.S. Department of Energy Office of Science.
\end{acknowledgments}

\bibliography{basename of .bib file}

\end{document}